\documentclass[10pt]{iopart}
\usepackage{graphicx}
\begin{document}
\title[Orbital moment of a single Co atom on a Pt-surface]{Orbital moment of a single Co atom
on a Pt(111) surface -  a view from correlated band theory.}

\author{Alexander B. Shick,$^1$ and
Alexander I. Lichtenstein$^2$}

\address{$^1$Institute of Physics ASCR, Na Slovance 2, Prague,
Czech Republic}
\address{$^2$University of Hamburg, Jungiusstrasse 9, 20355 Hamburg, Germany}
\ead{shick@fzu.cz}

\begin{abstract}
The orbital magnetic moment of a Co adatom on a Pt(111) surface is
calculated in  good agreement with experimental data making use of
the LSDA+U method. It is shown that both electron correlation
induced orbital polarization and  structural relaxation play
essential roles in orbital moment formation. The microscopic origins
of the orbital moment enhancement are discussed.
\end{abstract}
\maketitle

\section{Introduction.}
According to Hund's rules, gas-phase transition metal atoms possess
large spin $M_S$ and orbital $M_L$ moments mediated by intra-atomic
Coulomb interactions. In a solid, where electron delocalization and
crystal field effects compete with Coulomb interactions, there is a
substantial decrease in $M_S$ and partial or total quenching of
$M_L$.

Recent  X-ray magnetic circular dichroism (XMCD) measurements
\cite{gambardella03} report $M_S $ and $M_L$ of Co-adatom and small
Co clusters on Pt(111) in an ultra-high-vacuum. The $M_L$ = 1.1
$\pm$ 0.1 $\mu_B$ and $M_S + 7M_D$ = 1.8 $\pm$ 0.1 $\mu_B$ (where
$M_D$ is a spin dipole moment) were evaluated from XMCD spectra
using the conventional sum rules \cite{sumrules}. The number of
holes in the Co atom 3$d$-manifold $n_d$=2.4 was taken from local
spin-density calculations (LSDA). The XMCD experiments are
complimented by spin-polarized-relativistic Korringa- Kohn-Rostocker
(KKR) Green's function LSDA theoretical calculations, and $M_S$ =
2.14 $\mu_B$ and $M_L$=0.60 $\mu_B$ for Co site were obtained. No
structural relaxation of the Co atom position over the Pt surface
was considered, and the atomic-sphere approximation was employed.

The authors of Ref.\cite{gambardella03} assumed that the $M_L$
discrepancy between the KKR-LSDA theory and the XMCD experiments
originates from the lack of orbital polarization (OP) in LSDA. They
used the well known orbital polarization correction of Brooks
\cite{brooks}, adding to the LSDA total energy functional an {\em
ad-hoc} term $\frac{1}{2} B_R M_L^2$  with the LSDA calculated Racah
parameter $B_R$. This form of OP correction was widely used in the
past to improve upon $M_L$ in the bulk transition $d$- and
$f$-metals where sometimes it works well \cite{trygg95}.

For the Co-atom on Pt surface, Gambardella {\em et al.} found out
that the Brooks OP yields the $M_L$ which substantially exceeds the
experimental value. They had to reduce the LSDA calculated $B_R$ by
50 \% in order to obtain a $M_L$=1.50 $\mu_B$ comparable with the
experimental XMCD data. Ii is assumed in \cite{gambardella03} that
the reduction of $B_R$ compensates for a lack of structural
relaxation.

In this work we explore another avenue for the orbital polarization
correction to LSDA which is based on the correlated band theory
LSDA+U method \cite{LDAU}. It consists of LSDA augmented by a
correcting energy of a multiband Hubbard type and a
``double-counting" subtraction term which accounts approximately for
an electron-electron interaction energy already included in the
LSDA. Minimization of the LSDA+U functional generates not only the
ground state total energy, but also one-electron band structure
energies and spin-orbital states. The basic difference between
LSDA+U method and the LSDA is its explicit dependence on on-site
spin and orbitally resolved occupation matrices. The LSDA+U method
creates in addition to the spin-only dependent LSDA potential, the
spin and orbitally dependent on-site ``$+U$" potential which gives
OP beyond that given by the LSDA (where it comes from the spin-orbit
coupling only).

It was shown by Solovyev {\em et al.} \cite{igor98} that LSDA+U
produces the correct OP for insulating 3$d$-oxides.
Recent parameter-free GW calculations  for transition metal based
materials \cite{igor05} produce OP which is very similar to LSDA+U
results with the appropriate choice of Coulomb-$U$ \cite{shick2003}.

\section{Results and Discussion.}
We performed supercell calculations to model a Co adatom at a
Pt(111) surface. The supercell consists of three Pt(111) layers with
doubled (p(2x2)) 2-dimensional unit cell, and the Co atom on the top
taken in the {\em fcc} position (see Fig.~1).
The vacuum is modeled by the equivalent of two empty Pt layers. All
in-plane inter-atomic distances are adopted to be those of pure Pt.
The distance between the Co atom and the Pt surface was varied in
the calculations. We note that while the chosen supercell is quite
small, it provides separation of Co atoms beyond the second nearest
neighbors distance and includes interaction of Co with first and
second Pt nearest neighbors. Herein, we assume that the given
supercell is sufficient for Co $M_L$ calculations, which is mainly a
local quantity.

We use the LSDA+U method implemented in the full-potential
linearized augmented plane-wave (FP-LAPW) method including
spin-orbit coupling (SOC) \cite{shick99,shick04}. When SOC is taken
into account, the spin is no longer a good quantum number, and the
LSDA+U total-energy functional contains additional spin-off-diagonal
elements of the on-site occupation matrix $n_{m_1 \sigma_1,m_2
\sigma_2}$ \cite{igor98}. The LSDA contributions to the effective
potential (and corresponding terms in the total energy) are
corrected to exclude the  non-spherical interaction. It helps to
avoid the $d$-states non-spherical Coulomb and exchange energy
``double counting" of $d$-states in LSDA and ``+U" parts of the
effective potential and also corrects the  non-spherical
self-interaction of the $d$-states.

In the self-consistent calculations we used 48 special {\it
k}-points in combination with a Gaussian smearing for the {\it
k}-point weighting. A quasi-2D Brillioun zone (BZ) with $k_z =0$ was
adopted in order to simulate the 2D-character of the problem,
notwithstanding that the supercell calculations themselves are
inherently three dimensional. The ``muffin-tin'' radii used are
$R_{MT} =$2.2  a.u. for Co and 2.5 a.u. for Pt and $R^{Co}_{MT}
\times K_{max} =$7.7, with $K_{max}$ the cut-off for the LAPW basis.
The Coulomb-$U$=2 eV and exchange-$J$=0.9 eV were chosen which are
in the range of commonly accepted values for 3$d$-metals. In
principle, $U$ can be calculated by linear-response LSDA procedure
\cite{cococcioni04} or from GW \cite{igor05}, both yielding the
values $\sim$ 2 eV. As for exchange-$J$, it is not affected by
solid-state screening and equal to LSDA calculated Stoner exchange
parameter. The spin quantization axis is fixed along the
out-of-plane $z$-direction.

First let us make a comparison with the results of KKR for unrelaxed
geometry (see Table I.). The LSDA calculated values of $M_S$ and
$M_L$ agree quite well. However, the $M_L$ per $d$-hole is somewhat
bigger in KKR than in FP-LAPW calculations since $d$-shell
occupation is bigger in KKR (7.6) than in FP-LAPW (7.2). It is
probably due to the difference in the radius of ASA-spheres used in
KKR and MT-radius in FP-LAPW. Since in KKR calculations
\cite{gambardella03} they use the same radius for ASA-sphere for the
``big" Pt-atom and the ``smaller" Co-atom, this difference can
become significant and affect both the charge and spin density
distributions.

The XMCD experiments measure not the $M_L$ itself but the $M_L$ per
$d$-hole ratio $M_L/n_h$. The $M_L/n_h$ calculated in KKR and
FP-LAPW for unrelaxed geometry and making use of LSDA is a factor of
two smaller than one measured by XMCD. It was already mentioned
above that Ref.\cite{gambardella03} proposed the use of the Brooks
OP with reduced $B_R$ in order to improve the agreement with
experimental data. Here we show that the $B_R$ reduction alone does
not solve the problem. Rather it attempts to compensate for
limitations of  the calculations without taking account of
structural relaxation in open systems including an important class
of 3$d$-adatoms and clusters.

Next, we turn to the salient aspect of our investigation, the LSDA+U
calculations. When the Co-Pt inter-atomic distance as for pure Pt is
used, the calculated $M_L$ and $M_L/n_h$ are quite big (see Table
I.). By varying $d_{Co-Pt}$ and minimizing the total energy we find
the equilibrium $d_{Co-Pt} \approx$ 3.48 a.u., i.e. reduced by
almost 20 \%. Since we do not perform a full relaxation and the
amount of Pt in our supercell is relatively small, we can not claim
that this will be the correctly optimized $d_{Co-Pt}$. Nevertheless
it is reasonable to assume that calculated $d_{Co-Pt}$ distance is
approximately correct \cite{privcom}.

The change in $d_{Co-Pt}$ has strong effect on $M_L$ and $M_L/n_h$
(shown in Table I.). The $M_L/n_h$ becomes fairly close to the
experimental value and the agreement for $M_L$ is also substantially
improved. We should keep in mind that the ``experimental" value is
given as a product of measured $M_L/n_h$-ratio times the
KKR-calculated $n_h$ of 2.4. Making use of the LSDA+U calculated
$n_h$=2.9, we obtain the ``experimental" $M_L$ of 1.34 $\pm$ 0.12
$\mu_B$, which is in  good agreement with the LSDA+U calculated
value.

To understand how the enlargement of the Co $M_L$ in the LSDA+$U$
approach comes about we consider the spin and orbitally resolved
$3d$ densities of states ({\em d}DOS), which are shown in Fig.~2.
The spin-resolved {\em d}DOS (see Fig. 2(a)) reveals a substantial
narrowing of the band width from $\sim$6 eV for hcp Co to $\sim$4 eV
for the Co ad-atom as well as a moderate increase in the
spin-splitting, as is expected for the reduced Co coordination. The
spin-down DOS is split at the vicinity of $E_F$. When {\em d}DOS is
resolved in terms of cubic harmonics (see Fig.~2(b)), it becomes
clear that the spin-down peak below $E_F$ posses
$e_g:3z^2-r^2$-orbital character while the spin minority d-holes are
of $e_g:x^2-y^2$ and $t_{2g}$-orbital character.

Since the spin-up Co $d$-band is fully occupied, only changes of the
spin-down band are essential for the $M_L$ enhancement.  The
$m_l$-resolved Co-{\em d}DOS is shown in Fig.~2(c) for the spin-up
and spin-down channels. The major contribution to the increase of
$M_L$ originates from $| m_s=-\frac{1}{2} ; m_l=+2 \rangle$ orbital.
The $M_L$ enhancement is brought about by in-plane spin-down
${x^2-y^2}$ and $xy$ orbitals and much less affected by out-of-plane
${xz,yz}$ orbitals. The spin-down ${3z^2-r^2} (\sim |m_l=0\rangle)$
orbital does not contribute to $M_L$. This out-of-plane ${3z^2-r^2}$
orbital is the most localized due to the smallest overlap between
Co-3$d$ and Pt-5$d$ electrons.

It is necessary to mention that our analysis can not be regarded as
truly {\em ab-initio} due to the use of external Coulomb-$U$. Herein
we make use of a ``commonly used" value of $U$=2 eV while it can be
at least in principle obtained from constrained LSDA calculations
\cite{cococcioni04}. With increase of $U$ the $M_L$ value will
increase, and with decrease of $U$ it will decrease. Nevertheless
our results show quantitatively the role of Coulomb-$U$ in the $M_L$
formation.

Also, we did not consider here the magneto-crystalline anisotropy
(MAE) induced by Co-adatom. In contrast to $M_L$ which is mostly a
local property of the Co atom, the MAE will consist of contributions
from the Co atom as well as the Pt neighbors  due to strong Pt atom
SOC \cite{shick2003}. Most probably, the quantitative studies of the
MAE will require a bigger supercell; this is the subject of further
work.

Still we can make a rough estimate for the contribution of the Co
adatom into the MAE. When spin is rotated from the z-axis
(out-of-plane) to the x-axis (in-plane) direction there is only a
little change in the value of the Co atom $M_S$, from 2.14 $\mu_B$
(z-axis) to 2.16 $\mu_B$ (x-axis). The change in $M_L$ is
substantially greater, from 1.58 $\mu_B$ (z-axis) to 1.42 $\mu_B$
(x-axis). Indeed, this strong anisotropy in $M_L$ paves the way for
the strong MAE. Qualitatively, the Co atom contribution to the MAE
can be estimated making use of Bruno's relation \cite{bruno89}
$MAE[=(E_x - E_z)] \approx -\xi/4 (M_L^x -M_L^z)$, where $\xi$ is
the SOC constant (76 meV for Co-adatom). In the LSDA+U calculations,
we obtain the MAE of $\approx$ 3.2 meV/Co which is somewhat smaller
than the experimental value of $9.3 \pm 1.6$ meV. A similar estimate
for the LSDA calculations yields the MAE of ~2.0 meV/Co. While our
estimate gives the MAE which is smaller than the experimental data,
it is exceptionally large compared with other Co-based materials: a
few tenth of meV for Co/Pt and Co/Au multilayers, and 2.0 meV for Co
monatomic wire \cite{gambardella02}.

We note that the orbital moment enhancement has been recently
investigated in Ref. \cite{gambardella02} for the case of the Co
monatomic wire on the Pt(111) surface step edge. Making use of XMCD
the experimental value of the Co atom $M_L = 0.68 \pm 0.05 \; \mu_B$
was found , which is somewhat smaller than for the Co adatom case.
Also, it was shown that LSDA yields the Co monatomic wire $M_L \sim
0.16 \; \mu_B$ which is substantially smaller than the XMCD
experimental value (see e.g. Ref. \cite{baud06}). An account of
Coulomb-$U$ is increasing the $M_L$ value to 0.45 $\mu_B$
\cite{shick04}, improving substantially the agreement with the
experimental data. .

In conclusion, employing correlated band theory LSDA+U calculations
we have provided a microscopic picture of the anomalous enhancement
of the Co-adatom orbital moment. It is found that two major effects
need to be included in order to essentially improve the Co orbital
moment: (i) a correct LSDA+U orbital polarization due to the
Coulomb-$U$ and (ii) structural relaxation of the Co-Pt interatomic
distance. The calculated value of $M_L$ is found in fairly good
agreement with experimental XMCD data \cite{gambardella03} when
those effects are taken into account.

\section{Acknowledgments}
We gratefully acknowledge discussions with W. Wurth, P.M. Oppeneer,
and O.N. Mryasov. Financial support was provided by the Grant Agency
of the Academy of Sciences (Project A100100530), DFG Grant SFB668-A3
(Germany) and German-Czech collaboration program (Project
436TSE113/53/0-1, GACR 202/07/J047).

\newpage

\begin{table}
\caption{Spin ($M_s$),
 Orbital ($M_l$) magnetic moments (in $\mu_B$),
 and Orbital moment per $d$-band hole ($n_h$)  for a
 Co atom on Pt(111) resulting from the LSDA and LSDA+U calculations.}

\begin{tabular}{ccccccccc}
\hline \multicolumn{3}{c}{\bf Co ad-atom/Pt(111) }&
\multicolumn{2}{c}{KKR \cite{gambardella03}}&$M_S$ & $M_l$ & $M_l/n_h$ \\
\multicolumn{2}{c}{LSDA}&&&& 2.14   & 0.60     &  0.25 \\
\multicolumn{2}{c}{LSDA+OP/2}&&&& 2.14   & 1.50 & 0.63 \\
 \hline
\multicolumn{2}{c}{\bf CoPt$_{12}$}&&
\multicolumn{2}{c}{FP-LAPW}&$M_S$ & $M_l$ & $M_l/n_h$ \\
\multicolumn{2}{c}{LSDA}&&&& 2.18  & 0.57 &  0.20  \\
\multicolumn{2}{c}{LSDA+U}&\multicolumn{3}{c}{$U_{Co}$ = 2.0
eV, $J_{Co}$=0.9 eV} \\
\multicolumn{2}{c}{unrelaxed}&\multicolumn{2}{l}{$d_{Co-Pt}$=4.27 a.u.}&&2.23&2.07&0.70  \\
\multicolumn{2}{c}{relaxed}&\multicolumn{2}{l}{$d_{Co-Pt}$=3.48 a.u.}&&2.14&1.58&0.54 \\
\hline &\multicolumn{2}{l}{Experiment}&\multicolumn{3}{l}{{\bf XMCD
\cite{gambardella03}} $n_h = 2.4$}
  & 1.1$\pm$0.1 & 0.46$\pm$0.04\\
  &\multicolumn{2}{l}{Experiment}&\multicolumn{3}{l}{{\bf XMCD} ($n_h = 2.92$)}
   & 1.34$\pm$0.12 & \\
   \hline
\end{tabular}
\end{table}

\vspace*{3cm}

\begin{figure}[h]
\includegraphics[width=8cm,height=8cm]{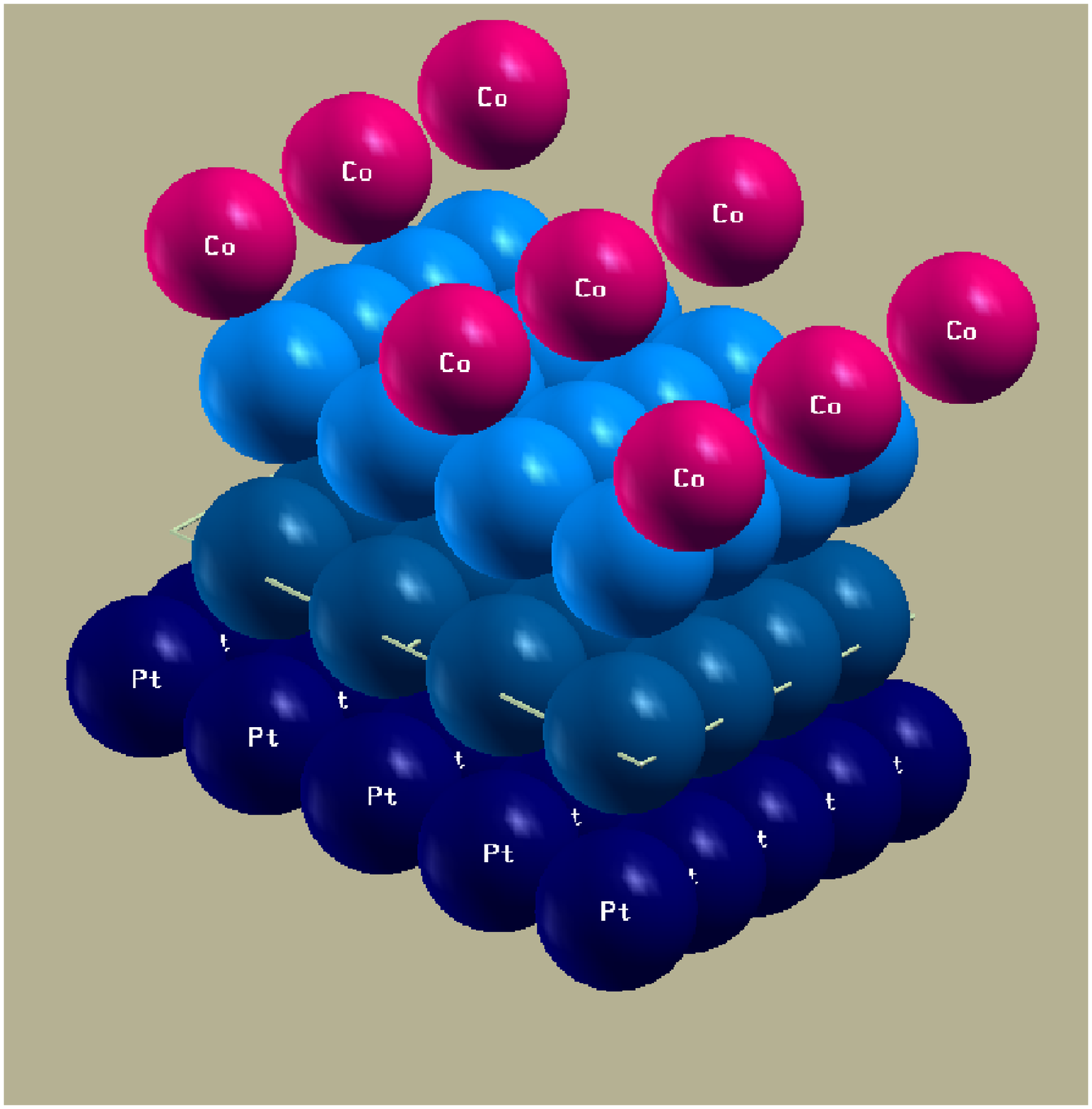}
\includegraphics[width=9cm,height=8cm]{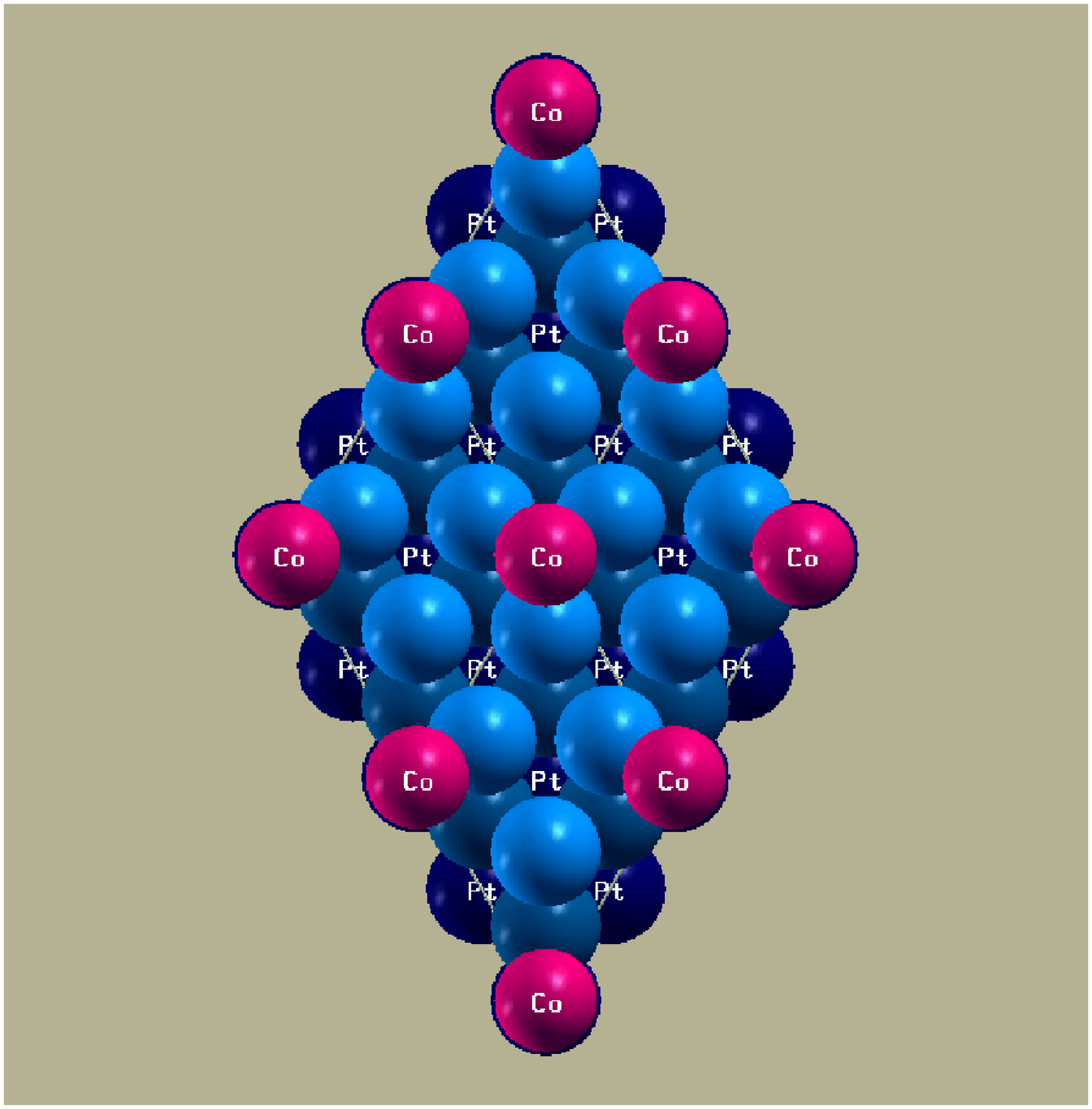}
\caption{Schematic crystal structure of a model super-cell: (left)
general view, (right) top-view with the Co-adatom in the {\em fcc}
position.} \label{fig1}
\end{figure}

\begin{figure*}[h]
\centerline{\includegraphics[width=10cm,height=10cm]{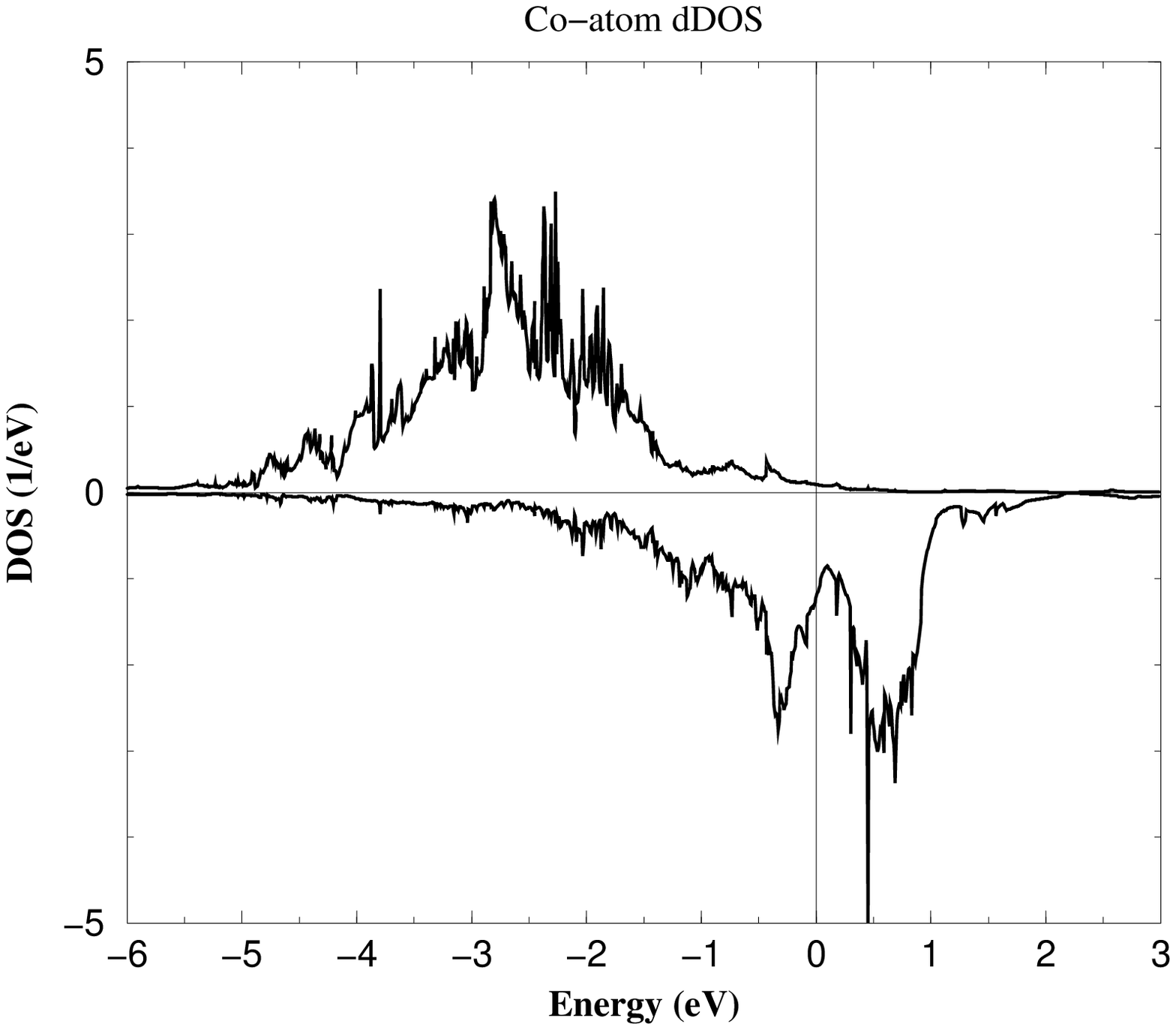}}
\centerline{\hspace*{2cm}
\includegraphics[width=10.5cm,height=10cm]{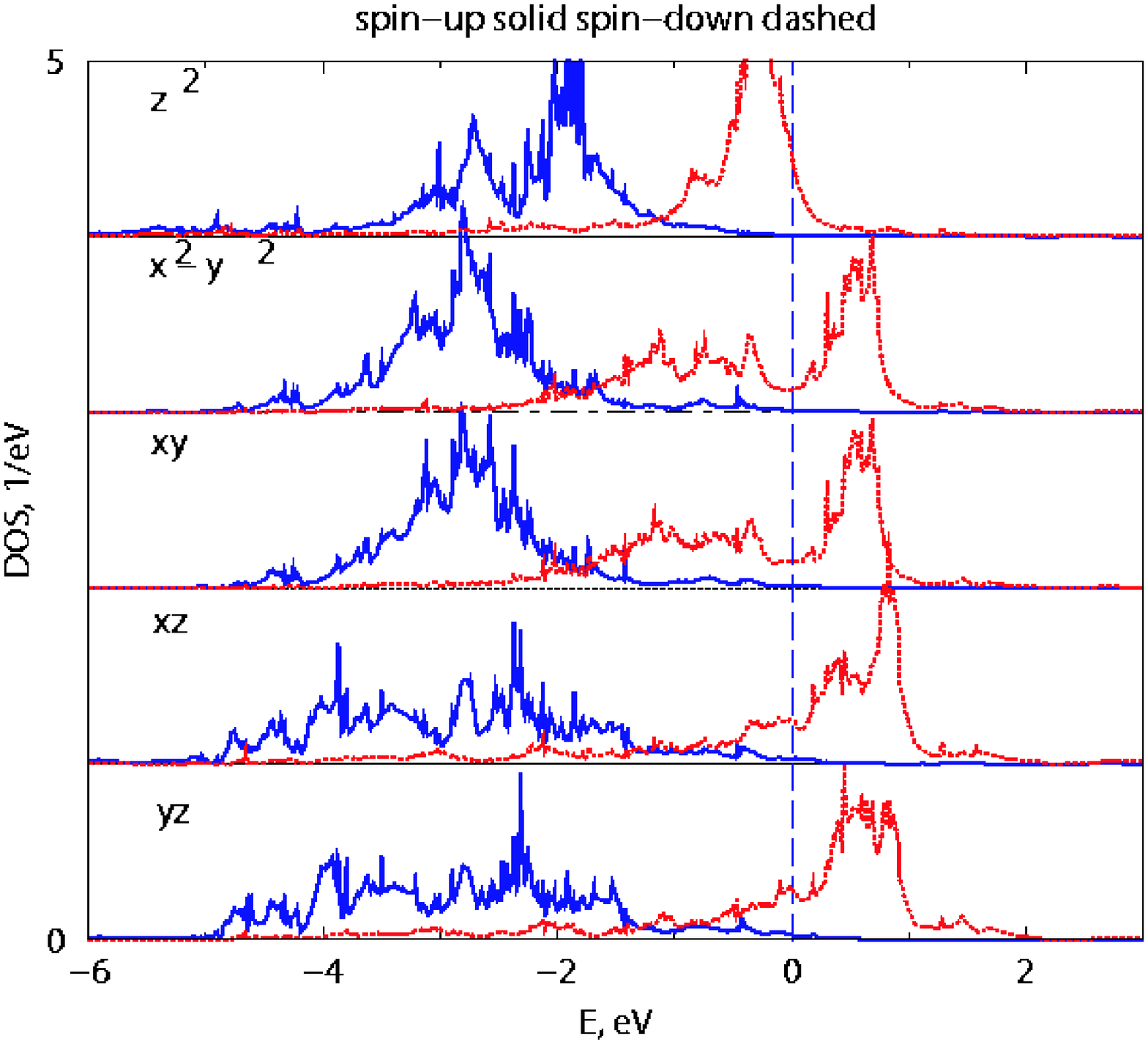}
\includegraphics[width=10.5cm,height=10cm]{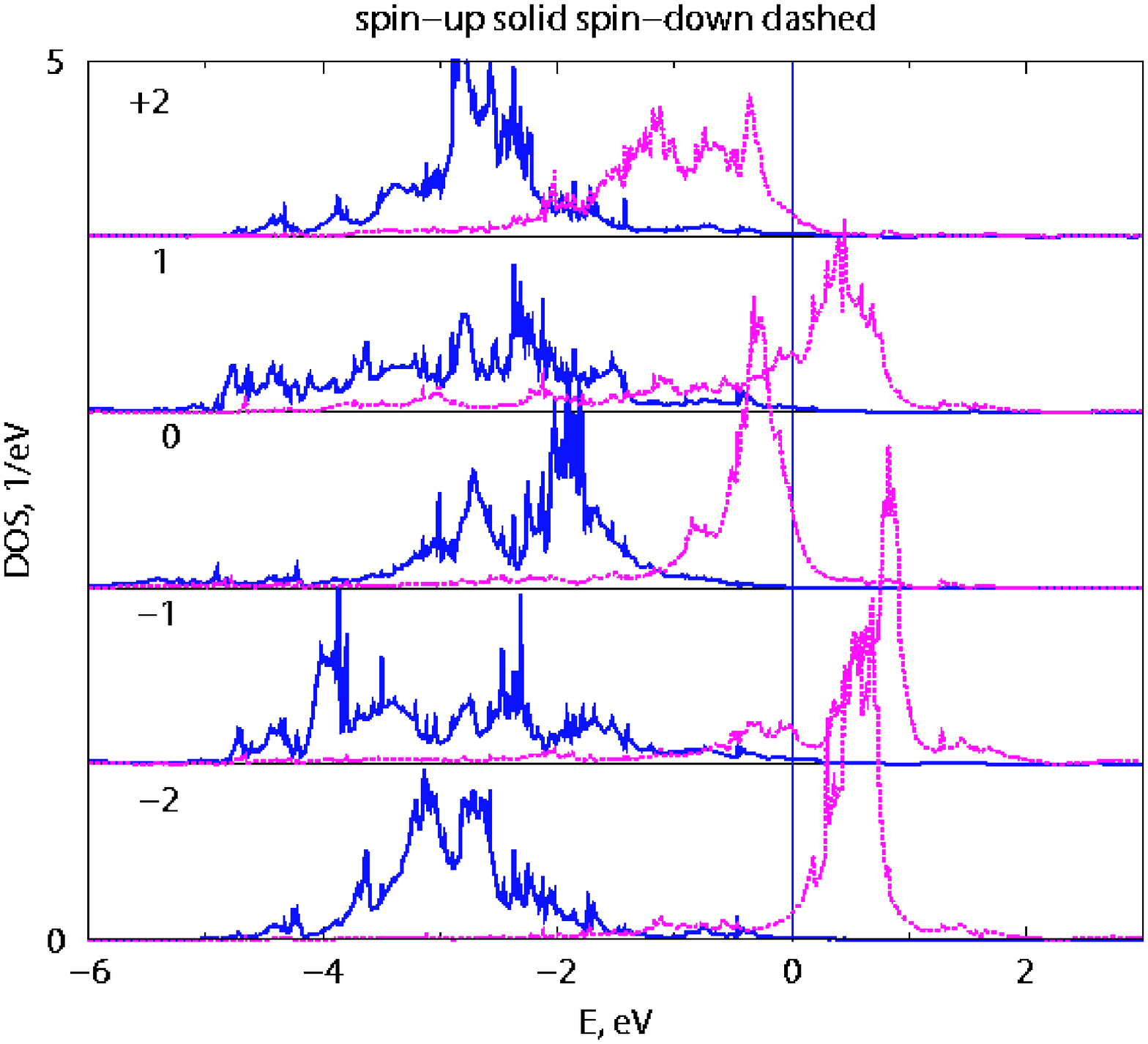}}
\caption{DOS for a Co adatom on Pt(111): (a,top) Spin-resolved
Co-atom {\em d}DOS; (b,left) Co-atom {\em d}DOS resolved in cubic
harmonics; (c,right) Co-atom {\em d}DOS resolved in complex
harmonics.}
\label{fig1}
\end{figure*}

\end{document}